\documentclass{article}
\usepackage{spconf,amsmath,graphicx}
\usepackage{xcolor}
\usepackage{amssymb}
\usepackage{caption}
\usepackage{float}
\title{Spec-NeRF: Multi-spectral Neural Radiance Fields}
\name{Jiabao~Li,
        Yuqi~Li$^*$,
        Ciliang~Sun,
        Chong~Wang,
        and~Jinhui~Xiang
        \thanks{This work is supported by Key Research and Development Program of Ningbo.}
        }
\address{Ningbo University}
\newcommand{\myname}{\textit{Spec-NeRF }}

\begin{document}
%
\maketitle
\begin{abstract}
We propose Multi-spectral Neural Radiance Fields(Spec-NeRF) for jointly reconstructing a multispectral radiance field and spectral sensitivity functions(SSFs) of the camera from a set of color images filtered by different filters. The proposed method focuses on modeling the physical imaging process, and applies the estimated SSFs and radiance field to synthesize novel views of multispectral scenes. In this method, the data acquisition requires only a low-cost trichromatic camera and several off-the-shelf color filters, making it more practical than using specialized 3D scanning and spectral imaging equipment. Our experiments on both synthetic and real scenario datasets demonstrate that utilizing filtered RGB images with learnable NeRF and SSFs can achieve high fidelity and promising spectral reconstruction while retaining the inherent capability of NeRF to comprehend geometric structures. Code is available at https://github.com/CPREgroup/SpecNeRF-v2.

\end{abstract}
\section{Introduction}
\vspace{-1em}

In high-fidelity scanning applications, there is a need to accurately reconstruct spectral and geometric information, as they provide insights into the inherent physical properties of objects. These attributes play a crucial role in representing the fundamental characteristics of the scanned objects, making their precise reconstruction highly desirable\cite{sun2023spectral}. Recently, there has been a growing fascination with merging 3D computer vision and spectral analysis. Numerous researchers have crafted spectral 3D models with the aim of enhancing the accuracy and dependability of computer vision tasks across diverse applications. These applications encompass areas like plant modeling \cite{liang20133d}, agriculture surveillance \cite{padua2019vineyard}, preservation of digital cultural heritage \cite{chane2013registration}, and material classification \cite{liang2014remote}.

The conventional methods can be classified into three categories \cite{sun2023spectral}.
\textit{1) Multi-source data fusion} includes mapping and estimation strategies, it combines different data types for better 3D models. 
The mapping strategy mainly focuses on projecting 2D spectral data onto a 3D point cloud \cite{elbahnasawy2018multi,lopez2022generation,graciano2020quadstack,lopez2021optimized,jurado2020multispectral} and the estimation strategy using active illumination to capture and estimate spectral reflection on the 3D structure \cite{li2019pro,li2021spectral}, 
especially, in \cite{li2021spectral}, multi-spectral data was acquired by directly using LED bulbs with various spectral power distributions.
To recover \textit{2) structures from spectra}, a standard Structure from Motion (SfM) technique is employed, generating 3D models band-by-band from multi-view images at the same wavelength, which are subsequently fused to create multispectral 3D models \cite{zia20153d,zia20163d}. 
\textit{3) Depth estimation}, spectral data can offer more depth cues (including reflectance, chromatic aberration, and defocus blur) than standard RGB images \cite{kumar2015defocus,ishihara2019depth,zia2015relative,zia2021exploring}. With a dual-camera system, \cite{heide2018real,luo2017augmenting} are able to obtain the pixels from multiple bands helping to get a better correlation for the stereo disparity.
In recent years, NeRF and its follow-up have made great strides in the field of multi-view synthesis in the field of 3D reconstruction. \cite{poggi2022cross} proposed the X-NeRF, given a set of images acquired from sensors with different light spectrum sensitivity (like infrared), which can learn a shared cross-spectral scene representation, allowing for novel view synthesis across spectra.

\begin{figure*}[htbp]
  \centering
  \includegraphics[width=\linewidth]{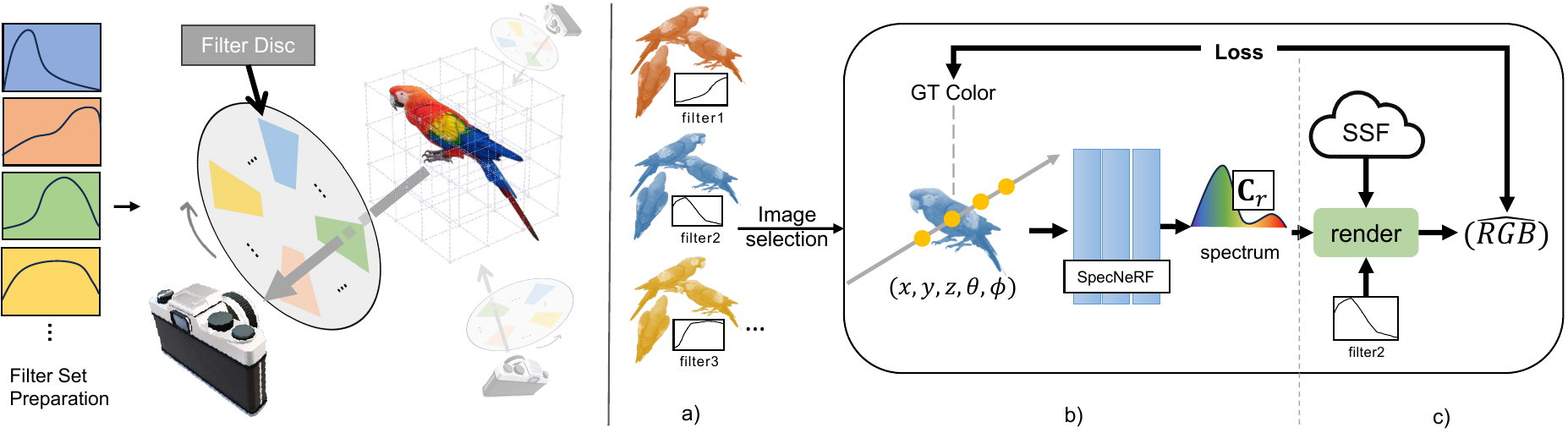}
  \vspace{-1em}
  \caption{The workflow of our proposed method, \myname. Left: The experimental setups. We utilize a filter wheel to switch the filters and acquire the RGB images of the scene using different filters at different positions. Right: The proposed Spec-NeRF network. a) The assembly groups of the training set, which are selected before sending to b), the key module, where a neural radiance field is adopted to reconstruct the spectral information. c) The spectrum is rendered into tristimulus values (e.g. RGB) using estimated SSF and the corresponding filter. The fidelity loss is built upon the RGB modality.}
  \vspace{-1em}
  \label{fig:framework}
\end{figure*}

In this paper, to provide the capacity of NeRF to recover the spectral information of a scene without the need for costly spectrum measuring devices and a priori knowledge of the Spectral Sensitive Function(SSF) of the used camera, we modify the conventional NeRF paradigm and propose the \myname, which only takes filtered RGB images from different views as input. By leveraging a NeRF framework, we simulate the filtered images using the known filter transmittance profiles and the estimated SSF, employing volume rendering techniques. This self-supervised learning approach enables us to learn the spectral characteristics of the scene effectively.
The contribution of this paper can be summarized as follows:
\begin{itemize}
\vspace{-1em}
    \item We present \myname to jointly recover the spectral and geometric information of the scene and the degradation parameter SSF. 
\vspace{-1em}
    \item Our method \textbf{only} leverages a regular camera and a few cheap filters, \textbf{without} known SSFs or constructing a training set, is able to achieve excellent spectra recovering results on both quantitative metrics and subjective evaluations.
\vspace{-1em}
    \item We construct both synthetic and real multi-view multi-spectral image datasets, providing a general-purpose benchmark for the training and evaluation of spectral 3D computer vision tasks.
\end{itemize}


\newcommand{\Yi}{\mathbf{Y}_i}
\newcommand{\Yihat}{\hat{\mathbf{Y}}_i}

\section{Method}

\subsection{Preliminary}
The NeRF network takes 5D coordinates $(x, y, z, \theta, \phi)$ of light rays as inputs and predicts the color and density $(\mathbf{c}, \sigma)$, which can be formulated as:
\begin{equation}
    \mathbf{c}, \sigma = F_\Theta (x, y, z, \theta, \phi)
\end{equation}
where $\Theta$ is the network parameters. Specifically, the color $\mathbf{C}_r$ of a camera ray, with $n$ sample points selected along the ray, is defined as follows:
\begin{equation}
    \mathbf{C}_r = \sum_{i=1}^{n} T_i \alpha_i \mathbf{c}_i
    \label{equ:renderC}
\end{equation}
where $T_i =  {\textstyle \prod_{j=0}^{i-1}}(1-\alpha_j)$ is the accumulated transmittance, also known as the probability of the ray casting ceases, $\alpha_i=1-\exp (-\delta_i \sigma_i)$ is the opacity.

\subsection{Spec-NeRF}
Our objective is to reconstruct a Neural Representation Field that encompasses the scene spectrum using $k$ spectral bands within a specific wavelength range (e.g. 380nm-730nm), and we can subsequently render a multi-spectral image from any desired novel viewpoint once the scene reconstruction is completed.
Hence, the NeRF network $F_\Theta$ outputs $\mathbf{c} \in \mathcal{R}^{1 \times k}$, and the rendered color $\mathbf{C}_r \in \mathcal{R}^{1 \times k}$ in Eq.\ref{equ:renderC} now represented as the spectrum.

To gather the spectral information of the scene, we capture the scene through distinct filter sets to RGB images, to do that we built an automatic capturing device to construct the real-world datasets presented in Fig.\ref{fig:framework} left, the camera and the rotatable filter disc are firmly bonded by a structure to ensure relative position remains unchanged, each filter will be rotated to the front of the lens before camera takes a shot.
Consider that the observed image has the size of $m \times n$, denoted as $\mathbf{Y} \in \mathcal{R}^{(mn) \times 3}$ (e.g., tristimulus color image) at each pose, is captured through a filter $\mathbf{f} \in \mathcal{R}^{1 \times k}$. This process can be regarded as a degradation model, expressed as follows:
\begin{equation}
    \small
    \Yi = (\mathbf{X}_{i} \circ \mathbf{f}) \mathbf{R}
    \label{equ:degradation}
\end{equation}
Here, $i \in [1, mn]$ represents the index of a specific pixel, while $\mathbf{X} \in \mathcal{R}^{(mn) \times k}$ denotes the multi-spectral image at a single viewpoint. $\mathbf{R} \in \mathcal{R}^{k \times 3}$ is denoted as the camera's SSF, $\circ$ donates the Hadamard product.

To recover the spectral and geometric information from the multi-view filtered images, we propose a modified NeRF framework called \myname, which adheres to the physical imaging processing described in Eq.\ref{equ:degradation} to learn the scene representation and estimate the SSF, as shown in Fig.\ref{fig:framework}.
Since the filter number is limited, and note that as the number of filters increases, the solutions to the equation become theoretically precise, hence, it is desirable to maintain a filter transmittance matrix with as high rank as possible, that is the filters should be distinguishable to each other and the total covering range of wavelength should be continuous.

To represent the camera parameter SSF $\mathbf{R}$ during the training of \myname, the most straightforward solution is to represent it directly as a matrix with $3k$ trainable parameters, or a neural function with positional encoding \cite{li2023busifusion}. However, this would introduce a greater degree of freedom to the optimization process. We address this problem by leveraging an implicit neural low-rank representation to represent the desired SSFs. The basis functions $\mathbf{b} \in \mathbb{R}^{k\times s}$ of SSFs can be extracted from SSFs datasets such as \cite{jiang2013space} by applying nonnegative PCA; and the coefficients $\mathbf{\alpha} \in \mathbb{R}^{s\times3}$ are generated by an MLP with positional encoding. Here $s$ denotes the amount of the basis functions. Therefore, the representation of SSFs is formulated as:
\begin{equation}
    \mathbf{R}  = \mathbf{b} \mathbf{\alpha}.
    \label{equ:eigenSSF}
\end{equation}

\subsection{Training}
Our experiment is conducted using the TensoRF framework \cite{chen2022tensorf}, however, it can be applied to other NeRF-based approaches, given its model-agnostic characteristics.
Now we define fidelity loss as 
\begin{equation}
    \mathcal{L}_{recon} = \frac{1}{2} \left \| \frac{\Yi - \Yihat}
    {sg(\Yihat) + \epsilon}  \right \|_F^2 \
\end{equation}
also we have $\Yihat = (\mathbf{C}_r \circ \mathbf{f_\mathbf{Y}}) \times \mathbf{R} $, in which $\mathbf{C}_r$ represents the volume-rendered spectrum, and $\mathbf{f}_\mathbf{Y}$ denotes the filter associated with the input image.
Note here we use relative MSE loss \cite{mildenhall2022nerf} to penalize the low radiance area, $sg()$ indicates a stop-gradient operator.
Furthermore, we utilize the distortion loss \cite{barron2022mip} to reduce the "floaters" in the empty space, but it is optional. To this end, the final training loss is:
\begin{equation}
    \mathcal{L} = \mathcal{L}_{recon} + \beta \mathcal{L}_{dist}
\end{equation}
where $\beta$ is a hyper-parameter. 

In real datasets, due to imperfections in the capturing procedure, thus, we initially train the network only using the images captured with the same filter for the first 2000 epochs to stabilize the geometry, then we adjust the learning rate to 1.2 times its initial value to avoid local minima and train the network with the rest images. Note that we do not adopt this strategy on synthetic datasets.

\begin{figure*}[ht]
  \centering
  \includegraphics[width=\linewidth]{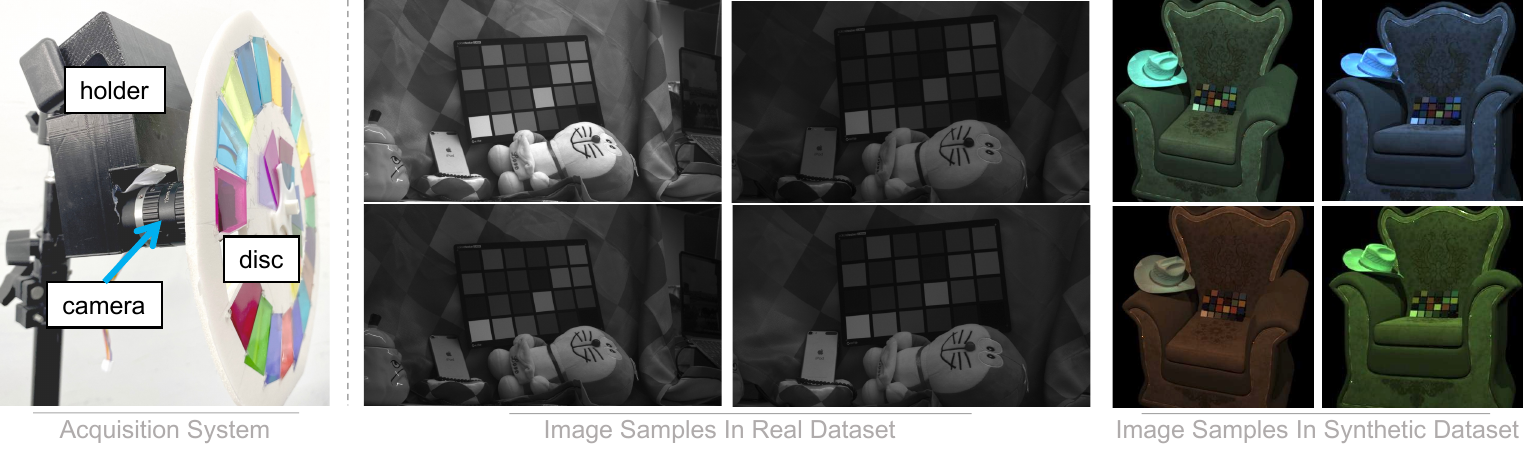}
  \vspace{-1em}
  \caption{A multi-view scenario capturing system for the image acquirement, and the resulting images in real datasets were filtered using the designated filters attached to the rotatable disc. Note that the camera utilized in our real dataset setup is a single-channel camera, capturing images in a monochromatic format, and a normal three-channel SSF is used in synthetic datasets.}
  \vspace{-1em}
  \label{fig:devices}
\end{figure*}

\section{Experiment}

\subsection{Experimental Setup \& Datasets}

Here we discuss the experimental settings and the instructions for constructing the real and synthetic datasets. 
We set $\epsilon$ to $0.01$, $\beta$ to $0.1$ if we employed the distortion loss term, and set spectral band number $k$ to 31 in the real dataset and 15 in the synthetic dataset. 
Our model is conducted on an NVIDIA 3090 GPU, with a batch size of 8192. The training process lasted for 25000 epochs, which took approximately 15 minutes to conclude when input consisted of 170 images.

\begin{figure}[H]
  \centering
  \includegraphics[width=\linewidth]{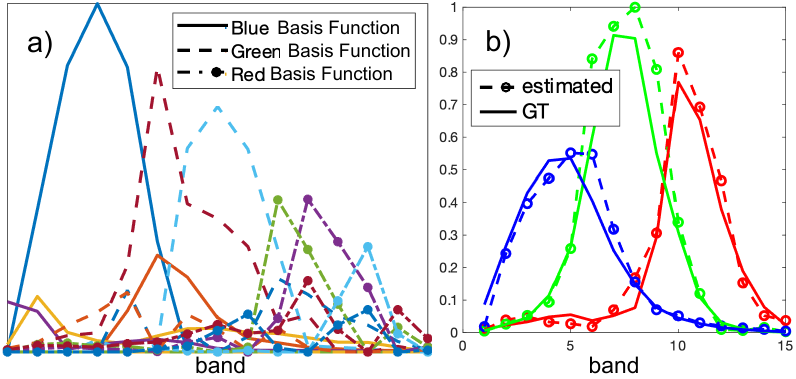}
  \caption{The visualization of the three channels basis matrices in a) derived from the decomposition of the NMF within the SSF database. In b), the jointly optimized three-channel SSF in a synthetic dataset using the basis function of NMF.}
  \label{fig:eigen}
\end{figure}

We specifically selected 25 SSFs from the database and identified the most plausible basis matrix through multiple iterations of the NMF algorithm, shown in Fig.\ref{fig:eigen} a). The rest three SSFs are for evaluation in synthetic dataset experiments.

\textit{For real scenario dataset}, images are captured with a single-band industrial camera, featuring an SSF with only one band, denoted as $\mathbf{R} \in \mathcal{R}^{k\times 1}$. The images have a resolution of $1920 \times 1080$ and RAW format is required for training.
During the scene capture process, we maintained constant camera settings such as the aperture size, ISO, and shutter speed. Specifically, we set the gain to $0$dB, and the gamma values were set to $(1, 1, 1)$. 
The transmittance profiles of the filters we measured cover the range of 430nm to 730nm with an interval of 10nm.

Apropos our capturing device, the disc with 20 uniformly spaced holes, allowing the filters to be securely attached. A stepping motor, which is connected to an ESP32 board, controls the rotation of the filter disc, and the camera and disc are combined using a 3D-printed holder, also a web interface was developed for controlling the camera's shutter when each filter rotates to the front of the lens.
The apparatus and multi-view images are presented in Fig.\ref{fig:devices}.

\textit{For synthetic dataset}, we built the scenes, especially a 24-color color checker, in Blender, which is then exported to the Mitsuba \cite{Jakob2020DrJit,mitsuba}. 
We used the "spectral" mode in Mitsuba, also adjusted the color checker's reflectance to match its true spectral profiles, after that the multi-spectral images were rendered with the size of $512 \times 512$ and covered the wavelength range from 440nm to 720nm with an interval of 20nm. Next, a synthetic filter set and a test SSF were adopted to generate the RGB images.

There remains redundancy in all the input images, hence we randomly select a specific proportion number of them, which means the input image has a different number and type of filters for each viewpoint.

\vspace{-2em}
\subsection{Results}
\vspace{-1em}

\subsubsection{Synthetic Scene}



\begin{figure}[ht]
  \centering
  \includegraphics[width=\linewidth]{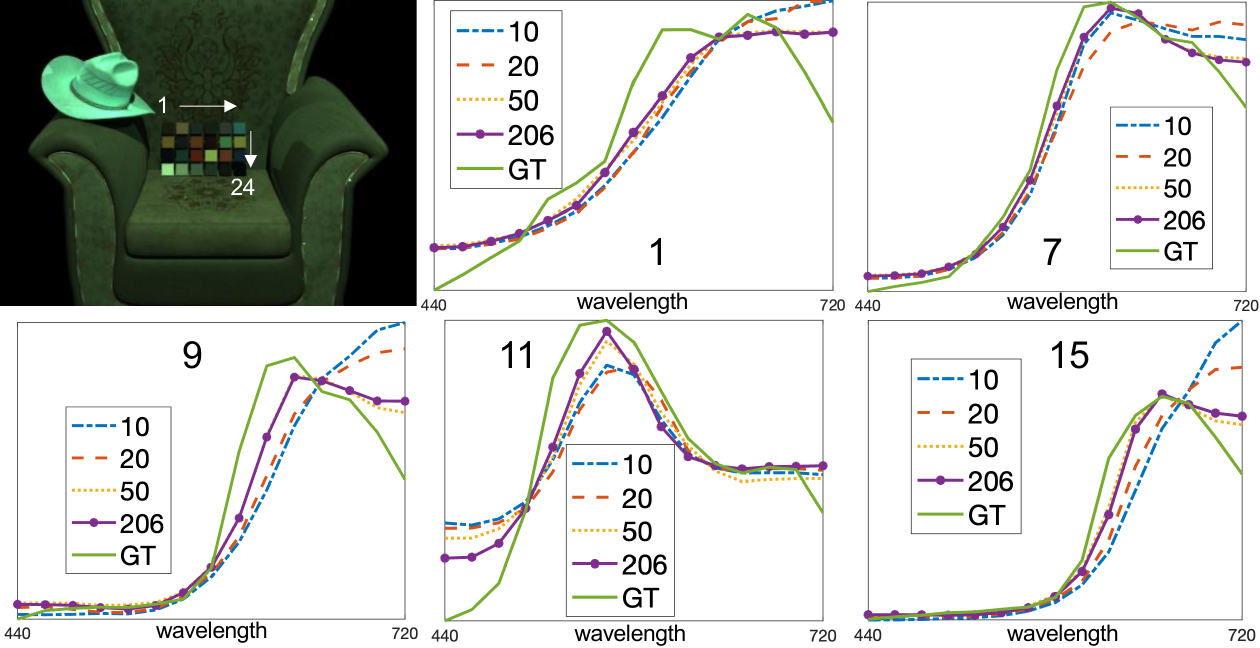}
  \caption{The reconstructed spectrum profiles of the color-checker block within the synthetic dataset. These profiles are generated using four different numbers of input images. The order of color blocks is from left to right, from top to bottom.}
  \label{fig:synRes}
\end{figure}

We first evaluate our model on a synthetic dataset and conduct a visual assessment of the estimated SSF and multi-spectral image.
Fig.\ref{fig:eigen} b) shows the estimated SSF by using the basis matrix of NMF, the experiment is performed by using 140 input images. 
Next, in Fig.\ref{fig:synRes} we present the reconstruction results of color-checker achieved by employing four different numbers of input images.
It exhibits the capability to approximate the ground truth shape even when the number of input samples is exceedingly low. As the number of input samples increases to a reasonable level, our approach consistently delivers high-quality results that closely align with the ground truth.

\subsubsection{Real Scene}
\vspace{-1em}

\begin{table}[]
\centering
\begin{tabular}{c|c|c|c}
\hline
PSNR  & SSIM  & LPIPS(alex) & LPIPS(vgg) \\ \hline
45.50 & 0.993 & 0.014       & 0.108      \\ \hline
\end{tabular}
\caption{Four metrics between the rendered and captured images on the real dataset.}
\label{tab:finalmetric}
\vspace{-1em}
\end{table}

\begin{figure}[ht]
  \centering
  \includegraphics[width=\linewidth]{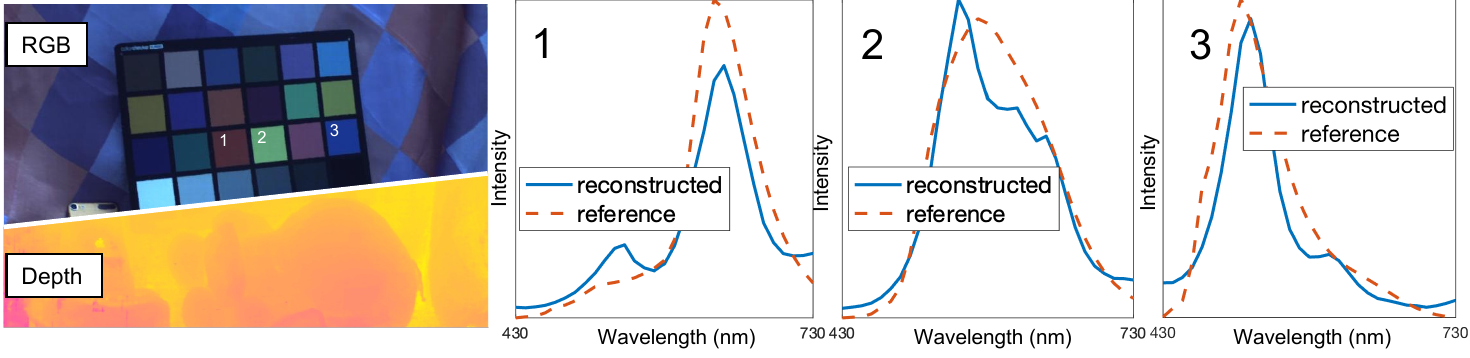}
  \caption{The rendered RGB image, depth map and spectrum reconstruction results on the real dataset at a novel view.}
  \label{fig:allresult}
\end{figure}

We further test our method on a real dataset captured by our apparatus in Fig.\ref{fig:devices}.
As the spectral reconstruction results of the checkerboard shown in Fig.\ref{fig:allresult}, 
the most representative three spectrums of color blocks (in color of R, G, B) are presented. Note that while we do not have the actual reflected spectrum of the color board, we possess the reflectance profiles of it from the manufacturer, and the ambient light spectrum is also measured, hence, with both, we can calculate a reflected spectrum as a "reference". 
The RGB image is rendered by employing an authentic SSF with three bands, and additionally, a depth map is rendered for geometry understanding visualization.

We present four quantitative metrics on the test set, as outlined in Tab.\ref{tab:finalmetric}, which have been computed on degraded images. 
These metrics compare two monochromatic images: one is directly captured, while the other is generated by rendering using the reconstructed multi-spectral image and the estimated SSF.

\begin{figure}[htbp]
  \centering
  \includegraphics[width=\linewidth]{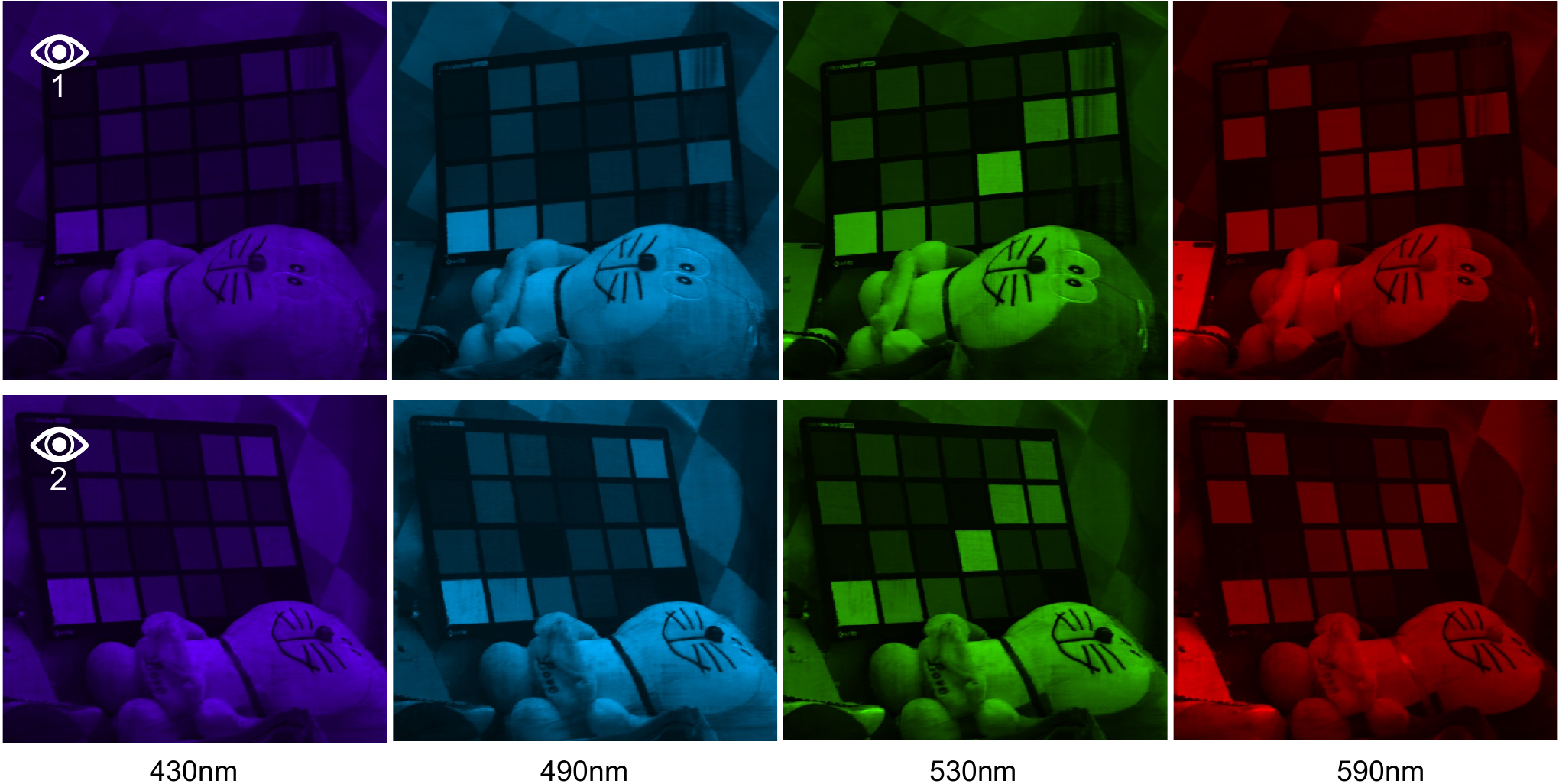}
  \caption{Visualization of the rendered multi-spectral images at two novel views, we present the scene at four specific wavelengths.}
  \label{fig:bandviz}
\end{figure}

Furthermore, we provide synthetic multi-spectral images from two novel viewpoints across four spectral bands in Fig.\ref{fig:bandviz}, demonstrating the capability of our method to generate high-fidelity images while preserving high resolution in both spatial and spectral dimensions.

\vspace{-3em}
\section{CONCLUSION}
\vspace{-1em}
In this paper, we propose \myname that jointly optimizes the degradation parameters and achieves high-quality multi-spectral image reconstruction results at novel views, which only requires a low-cost camera and filters.
We also provide two types of datasets for related studies.
For future work, we seek to decouple the albedo into ambient light and the reflectance of the surfaces with similar low-cost settings.


\bibliographystyle{IEEEbib}
\bibliography{refs}

\end{document}